\documentstyle[psfig,times]{jaa}
\begin{document}
\title[Multi-waveband Emission Maps of Blazars]{Multi-waveband Emission Maps of Blazars} 
\author[A.\ P.\ Marscher et al]%
       {Alan Marscher,$^1$\thanks{e-mail: marscher@bu.edu} Svetlana G. Jorstad,$^{1,2}$
        Valeri M.
        \newauthor
         Larionov,$^2$ Margo F. Aller,$^3$ and Anne L\"ahteenm\"aki$^4$ \\ 
        $^1$Inst. for Astrophysical Research, Boston U., 725 Commonwealth Ave., Boston,\\ MA 02215, USA\\
        $^2$Astronomical Inst., St. Petersburg State U., St. Petersburg, Russia\\
        $^3$Astronomy Dept., U. of Michigan, Ann Arbor, Michigan 48109-1042, USA\\
        $^4$Mets\"ahovi Radio Observatory at Aalto University, Kylm\"al\"a, Finland\\
}
\maketitle
\label{firstpage}
\begin{abstract}
We are leading a comprehensive multi-waveband monitoring program of 34 $\gamma$-ray bright blazars
designed to locate the emission regions of blazars from radio to $\gamma$-ray frequencies. The ``maps''
are anchored by sequences of images in both total and polarized intensity obtained with the VLBA
at an angular resolution of $\sim 0.1$ milliarcseconds. The time-variable linear polarization at radio
to optical wavelengths and radio to $\gamma$-ray light curves
allow us to specify the locations of flares relative to bright stationary features seen in the images
and to infer the geometry of the magnetic field in different regions of the jet. Our data reveal that
some flares occur simultaneously at different wavebands and others are only seen at some of the 
frequencies. The flares are often triggered by a superluminal knot passing through the
stationary ``core'' on the VLBA images. Other flares occur upstream or even parsecs downstream of
the core.
\end{abstract}

\section{Introduction}
\label{sec:intro}

One of the most fascinating properties of blazars is their ability to channel most of their apparent
luminosity into high-energy photons. The $\gamma$-ray luminosity can be as much as three orders of
magnitude higher than that at other wavebands. Furthermore, the emission regions must be very small,
since the time-scales of variability can be as short as hours. While relativistic beaming with
Doppler factors of 10-50 or even higher can bring down the luminosities and increase the upper
limits to the sizes of the emission regions, we are still faced with two problems: (1) getting TeV photons
out without rampant attenuation by pair production if they originate within $\sim 10^{17}$ cm of the
central engine, as many theorists favor, and (2) explaining the short time-scales of variability
if the $\gamma$-rays are emitted parsecs from the black hole. In order to determine which of these
difficulties we are facing, we need to specify where the $\gamma$-ray flares occur in blazars.
We are leading a collaboration that has mounted a comprehensive program aimed to do just that.

The heart of our program (see www.bu.edu/blazars)
is monitoring 34 $\gamma$-ray bright blazars $\sim$ monthly with the VLBA at 43 GHz, which
produces images of the compact jets with an angular resolution of $\sim 0.1$ milliarcseconds.
We also observe a subset of this sample, as well as some TeV blazars not in the sample, for 10-14 days
during intense campaigns two times per year in order to examine the short-term variability properties.
Every blazar contains
a bright feature at the upstream end that is referred to as the ``core.'' At lower frequencies,
the images contain a ``pseudo-core'' that is usually the location where the jet becomes optically
thick to synchrotron self-absorption. At millimeter wavelengths, however, the core in most
blazars has the properties of a physical feature with a quasi-stationary position. Since the
core is so prominent and is at least approximately stationary, it is very convenient to use as
a reference for locating the sites of flares at higher frequencies, where
images are far too coarse to resolve the nuclei.

Our technique for locating the flares relative to the core is most accurate when we can measure
the optical linear polarization of a flare and find that the electric vector position angle
$\chi$ is essentially the same as that of a feature seen (perhaps with a time delay) in the
VLBA images. If that value of $\chi$ is unique to the feature, then we can identify the flare
as occurring within that feature with a high level of confidence. If, as is very often the case,
the optical flare is coincident with a $\gamma$-ray and/or X-ray flare, then we can conclude
that the high-energy emission arises in the same feature. We can associate the flares based
solely on near-sumultaneity of peaks in the light curves as well {\it if}
such flares occur infrequently enough to render a chance coincidence highly unlikely.

Here we report the results of our monitoring program for some blazars that have exhibited
flaring events at $\gamma$-ray energies as observed by the {\it Fermi} Gamma-ray Space Telescope.
While some of the flares seem to occur between the 43 GHz core and the central engine, we find
that many events must occur at or beyond the core, which is located parsecs downstream of
the black hole. 

\section{Variety in the Behavior of Flares}
\label{sec:flares}

Figures 1 and 2 present multi-waveband light curves of BL~Lac and 3C~279, respectively. In the case
of BL~Lac, the X-ray, $\gamma$-ray, and radio fluxes all follow the same long-term rising trend starting
in early 2009, while the optical light curve does not do so until mid-2009. At the onset of this rise,
the X-ray flux increases by nearly a factor of 3 (Flare 1) over 40 days while the $\gamma$-ray emission
fluctuates, with a 2-day peak that is simultaneous
with the X-ray maximum. There is no sign, however, of the event
at the optical or radio bands. Flare 2 is a major, sharp $\gamma$/X-ray flare and a major radio outburst
that is delayed by one week, with only a hint of an optical counterpart. A superluminal knot
passes through the 43 GHz core on the VLBA images as the
radio flux rises. Another superluminal knot passes through the core near JD 2455090 during a more
pronounced optical and modest $\gamma$-ray flare (3), with no obvious sign of the event at the
X-ray energies. In early 2010, the very sharp Flare 4 occurs simultaneously at $\gamma$-ray, X-ray, and
optical frequencies as yet another knot crosses the core; a radio outburst starts at this
time but peaks about 3 weeks later. Flare 5, with very similar properties, peaks at JD 2455345.
Both flares 4 and 5 had precursors when the $\gamma$-ray flux undergoes rapid fluctuations.

\setcounter{figure}{0}
\begin{figure}
\begin{center}
\ \psfig{figure=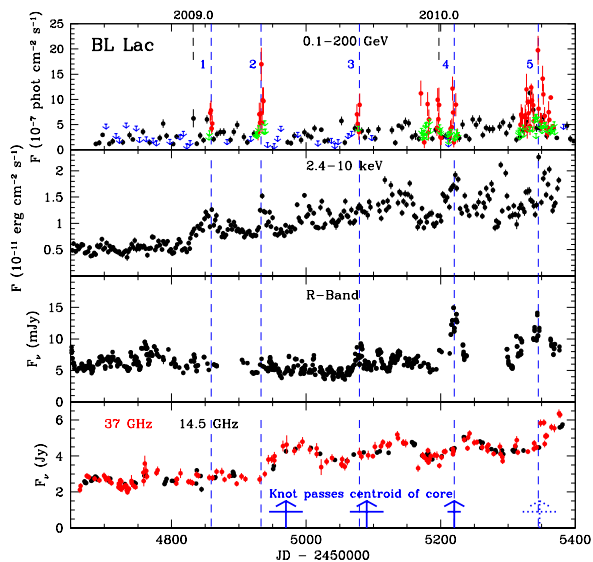,height=8cm}
\end{center}
\caption{Multi-waveband light curves of BL~Lac.  In the top panel, the data correspond to 5-day
integrations (black data points, blue 2-$\sigma$ upper limit arrows) or, during flares, 1-day integrations
(red data points, green upper limits); flares discussed in the text are marked by vertical blue
dashed lines. Times when superluminal knots passed
the core are marked with upward arrows, with the horizontal line segments indicating the
uncertainties in these times. The images in summer 2010 contain a knot whose motion has not yet
been determined; if its apparent speed is the same as for previous knots, it passes through the core
at the time indicated by the dotted arrow. From Marscher et al. (in prep.)}
\end{figure}

Each flare in 3C~279 seems to have its own multi-waveband behavior (see also Abdo et al. 2010).
The first, in late 2008,
includes multiple flares at both $\gamma$-ray and optical frequencies, with a new superluminal
knot passing through the core during the early stage of the outburst. The $\gamma$-ray light curve
during this period is dominated by two moderately sharp peaks, midway between which the flux drops
to about twice its typical quiescent level. After the second peak, the $\gamma$-ray and optical
emission becomes relatively quiescent. In the midst of this low state, a major X-ray outburst
occurs, accompanied by the passage of a superluminal knot across the 43 GHz core and a very minor
optical flare. Later in 2009, there is a double gamma-ray and optical flare with essentially
simultaneous peaks. The X-ray light curve contains a counterpart only for the second of these
flares, but the maximum in X-ray flux appears to lead the $\gamma$-ray peak by a few days. The
emission at all wavebands is relatively quiescent over the first eight months of 2010, during
which time no new bright knots appear in our 43 GHz VLBA images.

\setcounter{figure}{1}
\begin{figure}
\begin{center}
\ \psfig{figure=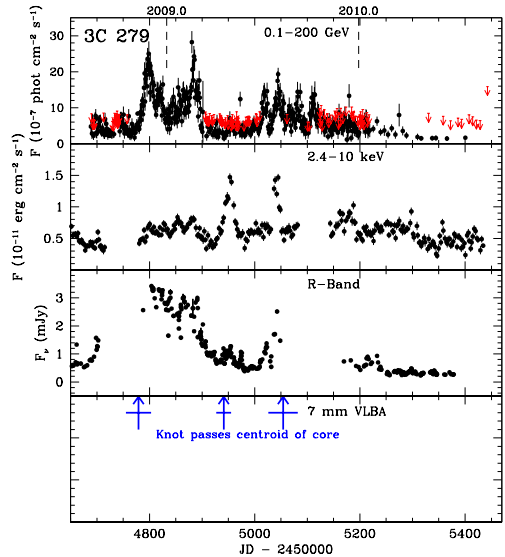,height=8cm}
\end{center}
\caption{Multi-waveband light curves of the quasar 3C~279. In the top panel, measurements (black)
and 2-$\sigma$ upper limits correspond to 1-day integrations. Blue arrows in bottom panel
have the same meaning as in Figure 1.}
\end{figure}

From the timing of the flares at different wavebands and the passage of superluminal knots across
the 43 GHz core, we infer that some, $\sim 50$\%, of the flares occur during such passages.
The others probably take place upstream of the core, although in OJ287 (Agudo et al. 2010)
and 3C~273 $\gamma$-ray flares
have been associated with a quasi-stationary feature $> 14$ pc {\it downstream} of the core
(Jorstad et al. 2010). Of particular interest in BL Lac are the
precursors to flares 4 and 5. These are probably similar to the first peak of the double flare
observed in 2005 (Marscher et al. 2008) and inferred to arise from the acceleration and collimation
zone of the jet. For that flare and another in PKS~1510$-$089, we observed rotations in the
optical linear polarization vector. The data are consistent with magnetic jet launching
models, with the jet plasma becoming turbulent beyond the point where the kinetic energy density
reaches equipartition with the magnetic field. 

The complex relationship among the light curves at the different bands defies the expectations of
simplistic one-zone or even basic multi-zone (colliding shock) models. Instead, turbulent
plasma blobs passing through standing shocks (e.g, in the core) may offer a better explanation for
the rather capricious behavior of the time variations in both flux and polarization (Marscher \&
Jorstad 2010).

\section{Conclusions}

By following changes in the 43 GHz images, linear polarization, and multi-waveband fluxes of
blazars, both during highly active and more quiescent states, we are able to specify the
locations of many flares. This, plus the correlations (or non-correlations) among wavebands,
provide numerous clues regarding the physics of the emission. We find that flares
can occur parsecs downstream of the central engine as well as upstream of the core, closer to the
black hole. Even for
events taking place parsecs down the jet, time-scales of variability can be shorter than
one day. There are various sources of seed photons, including the jet's own synchrotron radiation,
IR emission by hot dust, and an otherwise unobserved source, which may be synchrotron radiation
from a slower sheath of the jet. We cannot determine whether any of the flares occur close enough
to the central engine for the accretion disk or broad emission lines to contribute significantly
to the seed photon field.

As the {\it Fermi} mission continues, our program will sample a wider range of behaviors of
multi-waveband outbursts in blazars. Our data should keep theorists busy for many
years as they strive to solve the observational puzzles.

This research is supported in part by NASA grants NNX08AV65G, NNX08-AV61G, NNX09AT99G, NNX10AL13G,
NNX10AU15G, NNX09AU16G, \& NNX10AP16G, and National Science Foundation (NSF) grants
AST-0907893 \& AST-0607523.
The VLBA is an instrument of the National Radio Astronomy Observatory, a facility of the NSF,
operated under cooperative agreement by Associated Universities, Inc.


\begin{thebibliography}{10}
\bibitem{abdo:10} Abdo, A., et al., 2010, \newblock {\it Nature}, {\bf 463}, 919.
\bibitem{agudo:10} Agudo, I., et al., 2010, \newblock {\it Astrophys. J. Let.}, in press.
\bibitem{jorstad:10} Jorstad, S.G., et al., 2010, \newblock In Savolainen, T., et al., editors,
{\it Fermi Meets Jansky - AGN at Radio and Gamma-Rays}, Max-Planck-Institut f\"ur Radioastronomie, Bonn, 115.
\bibitem{marjor:10} Marscher, A.P., \& Jorstad, S.G., 2010, \newblock 
{\it ibid.}, 171.
\bibitem{marscher:08} Marscher, A.P., et al., 2008, \newblock {\it Nature}, {\bf 452}, 966.
\bibitem{marscher:10} Marscher, A.P., et al., 2010, \newblock {\it Astrophys. J. Let.},
{\bf 710}, L126.


\end{thebibliography}
\end{document}